\documentclass[%
 preprint,
preprint,
 amsmath,amssymb,
 aps, physrev,superscriptaddress
]{revtex4-2}

\usepackage{graphicx}
\usepackage{dcolumn}
\usepackage{bm}

\begin{document}

\preprint{APS/123-QED}

\title{\textbf{A New Low $Q^2$ Measurement of the Proton's $g_1$ Spin Structure Function from Longitudinal \& Transverse Polarized Data} 
}%

\newcommand\NSU{Norfolk State University, Norfolk, Virginia 23504, USA}
\newcommand\UNH{University of New Hampshire, Durham, New Hampshire 03824, USA}
\newcommand\UVA{University of Virginia, Charlottesville, Virginia 22903, USA}
\newcommand\WM{The College of William and Mary, Williamsburg, Virginia 23187, USA}
\newcommand\CAT{Catholic University of America, Washington, DC 20064, USA}
\newcommand\Jlab{Thomas Jefferson National Accelerator Facility,
  Newport News, Virginia 23606, USA }
\newcommand\Rutgers{  Rutgers University, New Brunswick, NJ 08901, USA}
\newcommand\MIT{Massachusetts Institute of Technology, Cambridge, MA 02139, USA}
\newcommand\XLA{Xavier University of Louisiana, New Orleans, LA 70125, USA}
\newcommand\Seoul{Seoul National University, Seoul 151-742, South Korea}
\newcommand\INFN{Istituto Nazionale di Fisica Nucleare, Sezione di Roma Tor Vergata, I-00173  Rome, Italy}
\newcommand\Kent{Kent State University, Kent, OH 44240, USA}
\newcommand\Catania{Universita di Catania, 95124 Catania, Italy}
\newcommand\USC{University of South Carolina, Columbia, SC 29208, USA}
\newcommand\Temple{Temple University, Philadelphia, Pennsylvania 19122, USA}
\newcommand\Argonne{Argonne National Laboratory, Argonne, Illinois 60439, USA}
\newcommand\CalLA{California State University, Los Angeles, Los Angeles, California 90032, USA}
\newcommand\Orsay{Institut de Physique Nucleaire, 91400 Orsay, France}
\newcommand\FIU{Florida International University, Miami, Florida 33199, USA}
\newcommand\USTC{University of Science and Technology, Hefei 230000, China}
\newcommand\CAS{Chinese Academy of Sciences, Beijing 100045, China}
\newcommand\HamptonU{Hampton University, Hampton, Virginia 23668, USA}
\newcommand\ODU{Old Dominion University, Norfolk, Virginia 23529, USA}
\newcommand\CNU{Christopher Newport University, Newport News, VA 23606, USA}
\newcommand\TelAviv{Tel Aviv University, Tel Aviv, 69978 Israel}
\newcommand\Yerevan{Yerevan Physics Institute, Yerevan, Armenia 0036}
\newcommand\Stefan{Jo\v{z}ef Stefan Institute, Ljubljana, Slovenia}
\newcommand\Hebrew{Hebrew University, Jerusalem 9190401, Israel}
\newcommand\Duke{Duke University, Durham, NC 27708, USA}
\newcommand\Glasgow{Glasgow University, Glasgow G12 8QQ, Scotland}
\newcommand\ORNL{Oak Ridge National Laboratory Oak Ridge, TN 37830, USA}
\newcommand\Kharkov{Kharkov Institute of Physics and Technology, Kharkov 61108, Ukraine}
\newcommand\Regina{University of Regina, Regina, Canada}
\newcommand\Mississippi{Mississippi State University, Mississippi State, MS
39762, USA}
\newcommand\Berkeley{Lawrence Berkeley National Laboratory, Berkeley, CA 94720, USA}
\newcommand\Ljubljana{Faculty of Mathematics and Physics, University of Ljubljana, Ljubljana, Slovenia}

\author{D.~Ruth}
  \email{Contact author: druth@nmsu.edu}

\affiliation{%
 New Mexico State University, Las Cruces, NM, USA 88003
}
\affiliation{\UNH}%

\author{K.~Slifer} \affiliation{\UNH}

\author{J.~P.~Chen} \affiliation{\Jlab}

\author{R.~Zielinski} \affiliation{\UNH}
\author{C.~Gu} \affiliation{\UVA}
\author{M.~Allada~(Cummings)} \affiliation{\WM}
\author{T.~Badman} \affiliation{\UNH}
\author{M.~Huang} \affiliation{\Duke}
\author{J.~Liu} \affiliation{\UVA}
\author{P.~Zhu} \affiliation{\USTC}
\author{K.~Allada} \affiliation{\MIT}

\author{A.~Camsonne} \affiliation{\Jlab}
\author{K.~Aniol} \affiliation{\CalLA}
\author{J.~Annand} \affiliation{\Glasgow}
\author{J.~Arrington} \affiliation{\Argonne} \affiliation{\Berkeley}
\author{T.~Averett} \affiliation{\WM}
\author{H.~Baghdasaryan} \affiliation{\UVA}
\author{V.~Bellini} \affiliation{\Catania}
\author{W.~Boeglin} \affiliation{\FIU}
\author{J.~Brock} \affiliation{\Jlab}
\author{C.~Carlin} \affiliation{\Jlab}
\author{C.~Chen}\affiliation{\HamptonU}
\author{E.~Cisbani} \affiliation{\INFN}
\author{D.~Crabb} \affiliation{\UVA}
\author{A.~Daniel} \affiliation{\UVA}
\author{D.~Day} \affiliation{\UVA}
\author{R.~Duve} \affiliation{\UVA}
\author{L.~El~Fassi} \affiliation{\Rutgers} \affiliation{\Mississippi}
\author{M.~Friedman} \affiliation{\Hebrew}
\author{E.~Fuchey} \affiliation{\Temple}
\author{H.~Gao} \affiliation{\Duke}
\author{R.~Gilman} \affiliation{\Rutgers}
\author{S.~Glamazdin} \affiliation{\Kharkov}
\author{P.~Gueye} \affiliation{\HamptonU}
\author{M.~Hafez} \affiliation{\ODU}
\author{Y.~Han} \affiliation{\HamptonU}
\author{O.~Hansen} \affiliation{\Jlab}
\author{M.~Hashemi Shabestari} \affiliation{\UVA}
\author{O.~Hen} \affiliation{\MIT}
\author{D.~Higinbotham} \affiliation{\Jlab}
\author{T.~Horn} \affiliation{\CAT}
\author{S.~Iqbal} \affiliation{\CalLA}
\author{E.~Jensen} \affiliation{\CNU}
\author{H.~Kang} \affiliation{\Seoul}
\author{C.~D.~Keith} \affiliation{\Jlab}
\author{A.~Kelleher} \affiliation{\MIT}
\author{D.~Keller} \affiliation{\UVA}
\author{H.~Khanal} \affiliation{\FIU}
\author{I.~Korover} \affiliation{\TelAviv}
\author{G.~Kumbartzki} \affiliation{\Rutgers}
\author{W.~Li} \affiliation{\Regina}
\author{J.~Lichtenstadt} \affiliation{\TelAviv}
\author{R.~Lindgren} \affiliation{\UVA}
\author{E.~Long} \affiliation{\UNH}
\author{S.~Malace} \affiliation{\USC}
\author{P.~Markowitz} \affiliation{\FIU}
\author{J.~Maxwell} \affiliation{\UNH} \affiliation{\Jlab}
\author{D.~M.~Meekins} \affiliation{\Jlab}
\author{Z.~E.~Meziani} \affiliation{\Temple}
\author{C.~McLean} \affiliation{\WM}
\author{R.~Michaels} \affiliation{\Jlab}
\author{M.~Mihovilovi\v{c}} \affiliation{\Ljubljana} \affiliation{\Stefan}
\author{N.~Muangma} \affiliation{\MIT}
\author{C.~Munoz Camacho} \affiliation{\Orsay}
\author{J.~Musson} \affiliation{\Jlab}
\author{K.~Myers} \affiliation{\Rutgers}
\author{Y.~Oh} \affiliation{\Seoul}
\author{M.~Pannunzio Carmignotto} \affiliation{\CAT}
\author{C.~Perdrisat} \affiliation{\WM}
\author{S.~Phillips} \affiliation{\UNH}
\author{E.~Piasetzky} \affiliation{\TelAviv}
\author{J.~Pierce} \affiliation{\Jlab} \affiliation{\ORNL}
\author{V.~Punjabi}\affiliation{\NSU}
\author{Y.~Qiang} \affiliation{\Jlab}
\author{P.~E.~Reimer}\affiliation{\Argonne}
\author{Y.~Roblin} \affiliation{\Jlab}
\author{G.~Ron} \affiliation{\Hebrew}
\author{O.~Rondon} \affiliation{\UVA}
\author{G.~Russo} \affiliation{\Catania}
\author{K.~Saenboonruang} \affiliation{\UVA}
\author{B.~Sawatzky} \affiliation{\Jlab}
\author{J.~Zhang} \affiliation{\Jlab}
\author{A.~Shahinyan} \affiliation{\Yerevan}
\author{R.~Shneor} \affiliation{\TelAviv}
\author{S.~\v{S}irca} \affiliation{\Ljubljana} \affiliation{\Stefan}
\author{J.~Sjoegren} \affiliation{\Glasgow}
\author{P.~Solvignon-Slifer} \affiliation{\UNH}
\author{N.~Sparveris} \affiliation{\Temple}
\author{V.~Sulkosky} \affiliation{\MIT}
\author{F.~Wesselmann} \affiliation{\XLA}
\author{W.~Yan} \affiliation{\USTC}
\author{H.~Yang} \affiliation{\CAS}
\author{H.~Yao} \affiliation{\WM}
\author{Z.~Ye} \affiliation{\UVA}
\author{M.~Yurov} \affiliation{\UVA}
\author{Y.~Zhang} \affiliation{\Rutgers}
\author{Y.~X.~Zhao} \affiliation{\USTC}
\author{X.~Zheng} \affiliation{\UVA}

\collaboration{E08-027 Collaboration}

\date{\today}

\begin{abstract}
The proton's spin structure has proven to be far more complicated than was originally believed, and has been the subject of a number of experimental investigations. 
Of particular interest are the spin structure functions $g_1$ and $g_2$, which can be used to generate moments to directly compare experimental results to Chiral Perturbation Theory and other theories of Quantum Chromodynamics (QCD). The proton's $g_1$ structure function has been the subject of two other recent low momentum transfer experiments, but there are currently no published low momentum transfer measurements which collected data on the proton structure functions using both a longitudinally-polarized and a transversely-polarized target at the same kinematics. In this paper, we present the longitudinally polarized results of the Jefferson Lab E08-027 experiment, along with linked moments which combine this new result with the previously published transversely-polarized data from the same experiment. These results provide a proton $g_1$ extraction measured with very high precision across the resonance region, and provide new information on the value of $g_1$ dependent sum rules and moments.
\end{abstract}

\maketitle


Despite huge amounts of experimental and theoretical work in the century since the discovery of the proton, the structure of the nucleon which accounts for the vast majority of the mass in the visible universe is still not well understood. In the high momentum transfer ($Q^2$) regime, where asymptotic freedom dominates, quantum chromodynamics (QCD) gives predictions for the behavior of the strong force and the proton's constituents that are governed by it~\cite{CHEN_2010}. In the low $Q^2$ regime, the cutting edge effective theories of QCD such as chiral perturbation theory ($\chi$PT) are often untested or in internal conflict over their predictions for many observables, a situation which can only be improved through comparison to nucleon structure data. A valuable test of $\chi$PT is a measurement of the spin structure function moments, quantities which describe the coherent, cumulative behavior of a nucleonic system of partons. These moments are formed from integrals of the spin structure functions $g_1$ and $g_2$. In this paper, we present for the first time the $g_1$ results of the Jefferson Lab E08-027 experiment, which measured both spin structure functions of the proton at very low $Q^2$. In a previous publication~\cite{g2p_nature} we presented our results for the $g_2$ structure function and the moments dominated by it, in this paper we focus instead on the E08-027 results for the $g_1$ structure function, which is driven by scattering off a longitudinally polarized target, and the moments which are usually considered to be dominated by $g_1$.

The E08-027 experiment ran in Jefferson Lab's Hall A in 2012, using an ammonia (NH$_3$) target spin-polarized with Dynamic Nuclear Polarization, and Jefferson Lab's spin-polarized electron beam with 85 nA of current and beam energies from 1.1 - 3.3 GeV~\cite{druth_thesis}. The experiment featured four kinematic settings with a target transversely polarized to the beam, and one kinematic setting with a target polarized longitudinally with respect to the beam; the data from this longitudinal setting is shown below. A septum magnet was used in conjunction with the Hall A High Resolution Spectrometers to obtain the small scattering angle necessary for these low $Q^2$ measurements. Additional details on the experimental procedure can be found in~\cite{g2p_nature}.

To extract the spin structure functions, a spin-polarized cross section difference was formed for each kinematic setting by comparing scattered electrons with either forward or backward helicity. These polarized cross section differences were formed by measuring double-spin asymmetries and combining these with the Bosted-Christy unpolarized cross section model~\cite{Bosted3}, as is described in detail in~\cite{g2p_nature} and calculated with:
\begin{equation}
    \Delta\sigma_{\parallel} = 2 A_{\parallel} \sigma_0
\end{equation}
\begin{equation}
    \Delta\sigma_{\perp} = 2 A_{\perp} \sigma_0
\end{equation}
where $A_{\parallel,\perp}$ is the asymmetry measured by E08-027 using a longitudinally polarized or transverse polarized target respectively, and $\sigma_0$ is the unpolarized cross section which in this case was generated with the Bosted-Christy model. The transverse polarized cross section difference $\Delta \sigma_\perp$ and longitudinal polarized cross section difference $\Delta \sigma_\parallel$ are related to the spin structure functions $g_1$ and $g_2$ by~\cite{KUHN20091}:
\begin{equation}
    \Delta\sigma_\parallel = \frac{4\alpha^2}{M\nu Q^2}\frac{E'}{E}\big[g_1(x,Q^2)\{E+E'\cos\theta\}-\frac{Q^2}{\nu}g_2(\nu,Q^2)\big]
\end{equation}
\begin{equation}
    \Delta\sigma_\perp = \frac{4\alpha^2}{M\nu Q^2}\frac{E'^2}{E}\big[\nu g_1(x,Q^2)+2 E g_2(\nu,Q^2)\big]
\end{equation}

Here $\alpha$ is the fine structure constant, M is the proton mass, $\nu$ is the energy transfer, and E and $E'$ are the incident and scattered electron energies. In general $\Delta \sigma_\parallel$ is dominated by $g_1$ and $\Delta \sigma_\perp$ is dominated by $g_2$. Because E08-027 collected both transverse and longitudinal data at a beam energy of 2.2 GeV (with a $Q^2$ range of 0.024-0.069 GeV$^2$, and an average $Q^2$ of 0.045 GeV$^2$), it is possible to extract both $g_1$ and $g_2$ from the above equations without using a Spin-Structure-Function Model. A correction was applied to adjust the measured $g_1$ to a constant $Q^2$ of 0.045 GeV$^2$ using the Hall B Model~\cite{EG1b2}, see~\cite{g2p_nature} for more details on this procedure, which contributes a $<$2\% adjustment to the structure function that is included as a systematic uncertainty. This analysis method was used to produce the results shown below.
\begin{figure}
    \centering
    \includegraphics[trim={5cm 0.5cm 5cm 1cm},width=0.7\linewidth]{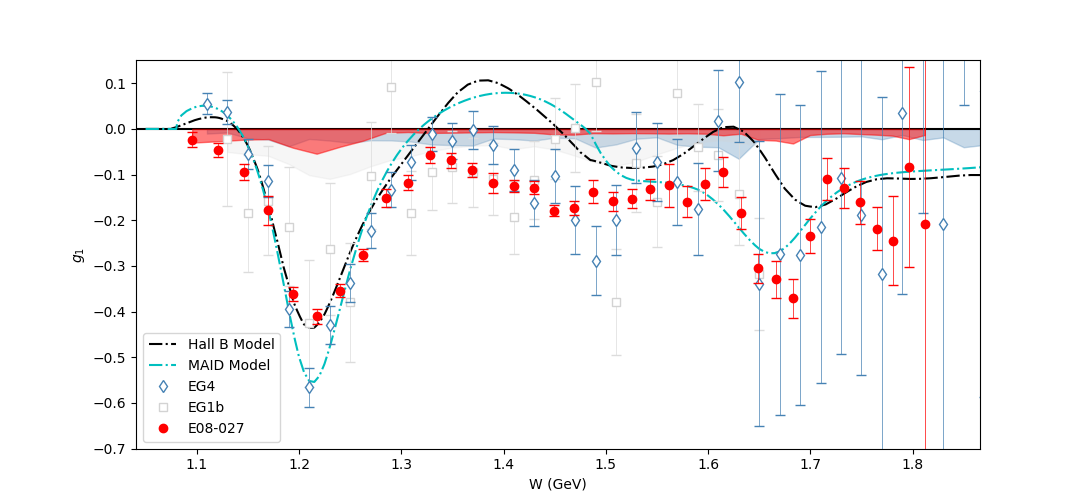}
    \caption{The $g_1$ proton structure function results of E08-027, at an average $Q^2$ of 0.045 GeV$^2$. Data is shown with red circles and statistical uncertainty in red error bars. The filled region indicates the systematic uncertainty. The black and cyan dash-dot lines represent the phenomenological Hall B ~\cite{EG1b2} and MAID~\cite{MAID2007} models respectively. The data is compared to the results from EG4~\cite{EG4_longpaper} at a $Q^2$ of 0.041 GeV$^2$, and from EG1b~\cite{EG1b2} at a $Q^2$ of 0.05 GeV$^2$.}
    \label{fig:g1}
\end{figure}

Our $g_1$ results are shown in Fig.~\ref{fig:g1}. Because we measured both longitudinally and transversely polarized proton asymmetries at an average $Q^2$ of 0.045 GeV$^2$, the displayed structure function does not require any $g_{1,2}$ or $A_{1,2}$ model as an input, which is a unique advantage of this new result for the proton. The data from E08-027 cover all the way down to the pion production threshold (W = 1078.2 MeV) with very good resolution thanks to the precision energy resolution of the Hall A spectrometers. The phenomenological models and other experimental data in this region show the structure function briefly crossing zero and becoming positive just above the pion production threshold, while our data shows different behavior and remains negative (though consistent within uncertainties with a positive result). The value at threshold is crucial to the higher moments (such as the $\gamma_0$ polarizability discussed below) due to a kinematic weighting which causes these lowest W data points to contribute either a strong positive or strong negative contribution to the overall integral, see Equation~\ref{eq:gamma0}. This can also have significant impact on the $\chi$PT calculations of hydrogen hyperfine splitting and lattice QCD extrapolations. The source of this tension between our data and other measurements and models near the pion production threshold is not yet well understood. 

In comparison with data from the EG4 experiment~\cite{EG4_longpaper}, which ran in Jefferson Lab's Hall B at similar $Q^2$, the resonances displayed in the $g_1$ spectrum of the E08-027 data are similar but the $\Delta$(1232) resonance is smaller in magnitude. The E08-027 $\Delta$(1232) resonance in particular is less negative than the result from EG4, agreeing more closely with the EG1b result~\cite{EG1b2}, while the EG4 result agrees better with the MAID phenomenological model~\cite{MAID2007}. 
E08-027 collected very high precision $g_1$ data at one $Q^2$ point and over an invariant mass of 1.078 - 1.8 GeV. 
The integrated moments of $g_1$ from E08-027 are shown and discussed below, with numerical values provided in Table~\ref{tab:moments}.

\begin{table}
\begin{center}
\begin{tabular}{||c | c | c | c||} 
 \hline
  & Value & Stat. & Syst. \\ [0.5ex] 
 \hline\hline
 $\Gamma_1^p$ & -0.0171 & 0.0005 & 0.0024 \\ 
 \hline
 $I_A^p$ & -0.954 & 0.0381 & 0.134 \\
 \hline
 $\gamma_0^p$ & -2.249 & 0.160 & 0.315 \\ [1ex] 
 \hline
\end{tabular}
\caption{The proton moment results of E08-027 at a $Q^2$ of 0.045 GeV$^2$. Each entry lists the central value $\pm$ the statistical uncertainty $\pm$ the systematic uncertainty. \label{tab:moments}}
\end{center}
\end{table}

\begin{figure}
    \centering
    \includegraphics[width=0.9\linewidth]{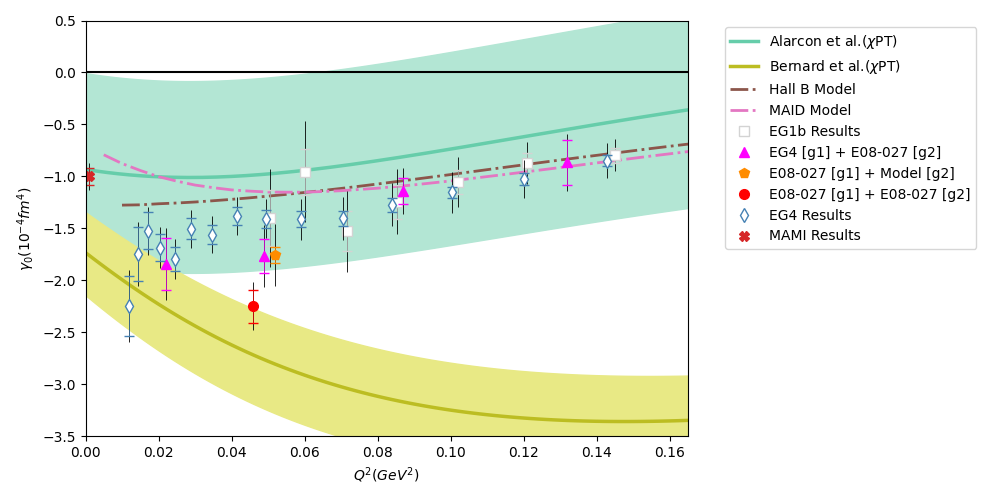}
    \caption{The $\gamma_0$ proton moment results of E08-027. The results from this paper are shown in the solid red circle, alongside results from EG4~\cite{EG4_final} in open blue diamonds and EG1b~\cite{EG1b2} in open grey squares. The EG1b points are offset slightly by 0.001 GeV$^2$ to allow comparison of error bars with EG4. The magenta triangles indicate results calculated using a combination of $g_2$ from E08-027 and $g_1$ from EG4 (adjusted to match the $Q^2$ of E08-027 using the Hall B model). The orange pentagon is the result calculated using an integral including the $g_1$ of E08-027 along with a $g_2$ produced with the Hall B model~\cite{EG1b2}. This orange point still includes a contribution from E08-027's transverse data in the extraction of $g_1$. The magenta and orange points at $Q^2$=0.046 GeV$^2$ are offset slightly by 0.003 and 0.006 GeV$^2$ respectively to allow visual comparison of the error bars. The red X indicates MAMI photoproduction results at $Q^2$=0~\cite{MAMI_gamma0}. The filled regions represent calculations from Chiral Perturbation Theory~\cite{Gold,Krebs}. The dash-dot lines indicate phenomonological models\cite{EG1b2,MAID2007}.}
    \label{fig:gamma0}
\end{figure}

For this work, the moment of greatest interest is the generalized forward spin polarizability $\gamma_0$~\cite{Gold}. The polarizabilities are a class of observables which measure a nucleon's ensemble response to an external field. $\gamma_0$ tracks a spin-dependent response, and is highly important for its capability to directly test effective theory calculations, of which there are currently two cutting-edge $\chi$PT calculations by different groups~\cite{Gold,Krebs}. $\gamma_0$ is written in terms of the spin structure functions as:

\begin{equation}
\label{eq:gamma0}
\gamma_0 = \frac{16 \alpha M^2}{Q^6} \int_0^{x_{\pi}} x^2 \big[g_1(x,Q^2) - \frac{4 M^2 x^2}{Q^2} g_2(x,Q^2)\big]dx 
\end{equation}

Here, $x_\pi$ is the Bjorken-x of the pion production threshold. $\gamma_0$ is often considered to be mostly determined by $g_1$ due to the kinematic weighting of the integral, but at low $Q^2$ and large $x$, the $g_2$ part becomes just as important. $\gamma_0$ is also strongly weighted towards spin structure function values near the pion production threshold due to the $x^2$ and $x^4$ weighting. Our $\gamma_0$ result is shown in figure~\ref{fig:gamma0}. There is an interesting and noticeable difference from the Hall B results of EG4 and EG1b~\cite{EG4_final,EG1b}. Both of these experiments favor the Alarcon et al.~\cite{Gold} prediction at a similar $Q^2$ to our data, and cross to favor the Bernard et al.~\cite{Bernard} calculation at low $Q^2$. Our result at $Q^2$ = 0.045 GeV$^2$ favors the Bernard et al. calculation. 

The difference in $\gamma_0$ between our result and the Hall B data flows largely from two places: firstly, as mentioned before, the moment is strongly weighted towards the pion production threshold region at large Bjorken x, where our data for $g_1$ are negative but the Hall B data are positive. Secondly, we include data from longitudinal- and transverse-polarized proton scattering~\cite{g2p_nature}, while the Hall B result is constructed only from ~\cite{EG4_final} longitudinal data. Though the differences for $g_2$ between E08-027 data and the Hall B model appear small~\cite{g2p_nature}, the $g_2$ part of the $\gamma_0$ integral has a very strong $x^4$ weighting. Consequently, because our $g_2$ data near the pion production threshold is slightly more positive than the model at $Q^2$=0.045 GeV$^2$, there is a sizable negative contribution to $\gamma_0$ as compared to the value using the spin structure function model for $g_2$. This is shown in the difference between the red and orange points in Figure~\ref{fig:gamma0}.


The impact of the E08-027 data for the $g_1$ and $g_2$ part of the integral can be seen separately in the magenta and orange points in Figure~\ref{fig:gamma0}. The magenta points use the E08-027 $g_2$ results for four kinematic settings combined with a $g_1$ at a similar $Q^2$ from the EG4 experiment~\cite{EG4_final}. These points are largely in good agreement with the Hall B data, but are slightly more negative at lower $Q^2$ because of the contribution from E08-027's transverse data. The orange point represents a $\gamma_0$ formed with the E08-027 $g_1$ and the Hall B model for $g_2$. This point is in the closest agreement with the Hall B data, but shows a negative shift due to the previously discussed sign difference at the pion production threshold. These two discrepancies combine to form the larger difference between the EG4 result and the red point formed solely from E08-027 data.


\begin{figure}
    \centering
    \includegraphics[width=0.9\linewidth]{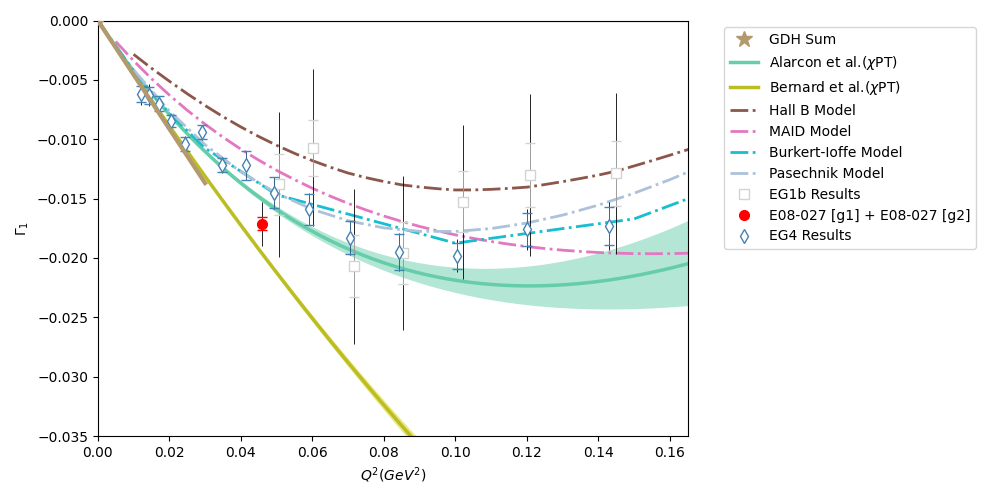}
    \caption{The $\Gamma_1$ proton moment results of E08-027. The results from this paper are shown in the solid red circle, alongside results from EG4~\cite{EG4_final} in open blue diamonds and EG1b~\cite{EG1b2} in open grey squares. The EG1b points are offset slightly by 0.001 GeV$^2$ to allow comparison of error bars with EG4. The filled regions represent calculations from Chiral Perturbation Theory~\cite{Gold,Krebs}. The dash-dot lines indicate phenomonological models\cite{EG1b2,MAID2007,pasechnik,burkert-ioffe}. The cyan star and line represent the Gerasimov-Drell-Hearn (GDH) slope~\cite{GDH}.}
    \label{fig:Gamma1}
\end{figure}

Another important moment of $g_1$ is its first moment, referred to as $\Gamma_1$, and can be calculated with:

\begin{equation}
\Gamma_1 = \int_0^1 g_1(x,Q^2) dx
\end{equation}

Where $x$ is the Bjorken variable $x = \frac{Q^2}{2 M \nu}$. $\Gamma_1$ is a uniquely useful moment for investigating nucleon structure, and is constrained at low $Q^2$ by the Gerasimov-Drell-Hearn Sum rule, which states~\cite{GDH}:

\begin{equation}
\label{eq:GDH}
\lim_{Q^2\to0}\frac{2 M^2}{Q^2}\Gamma_1(Q^2) = -\frac{\kappa^2}{4}
\end{equation}
with $\kappa$ as the anomalous magnetic moment of the nucleon. The slope needed to approach this value as $Q^2\to0$ is known as the GDH slope. Our results for $\Gamma_1$ are shown in figure~\ref{fig:Gamma1}. Because the $\Gamma_1$ integral extends all the way to $x$ = 0, it is necessary to fill in the low-$x$ part of the moment using the CLAS Hall B Model~\cite{EG1b2}. 

\begin{figure}
    \centering
    \includegraphics[width=0.9\linewidth]{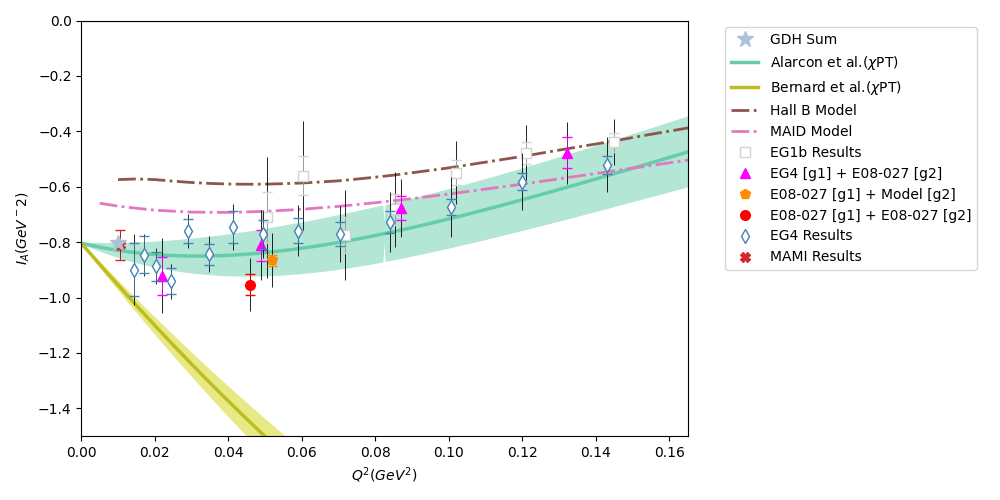}
    \caption{The $I_A$ proton moment results of E08-027. The results from this paper are shown in the solid red circle, alongside results from EG4~\cite{EG4_final} in open blue diamonds and EG1b~\cite{EG1b2} in open grey squares. The EG1b points are offset slightly by 0.001 GeV$^2$ to allow comparison of error bars with EG4. The magenta triangles indicate results calculated using a combination of $g_2$ from E08-027 and $g_1$ from EG4 (adjusted to match the $Q^2$ of E08-027 using the Hall B model). The orange pentagon is the result calculated using the $g_1$ of E08-027 along with a $g_2$ produced with the Hall B model~\cite{EG1b2}. The magenta and orange points at $Q^2$=0.046 GeV$^2$ are offset slightly by 0.003 and 0.006 GeV$^2$ respectively to allow visual comparison of the error bars. The filled regions represent calculations from Chiral Perturbation Theory~\cite{Gold,Krebs}. The dash-dot lines indicate phenomonological models\cite{EG1b2,MAID2007,Simula:2001iy}.}
    \label{fig:IA}
\end{figure}

Finally, we can investigate the generalized GDH integral $I_A$:

\begin{equation}
I_A = \frac{2 M^2}{Q^2} \int_0^{x_{\pi}} \big[g_1(x,Q^2) - \frac{4 M^2 x^2}{Q^2} g_2(x,Q^2)\big]dx 
\end{equation}

The result for this is shown in Figure~\ref{fig:IA} and is meant to approach the GDH result of equation ~\ref{eq:GDH} as the moment approaches $Q^2 = 0$. Here, the E08-027 data is in close agreement with the results of EG4 and EG1b from Hall B. The data are also near in magnitude to the $\chi$PT calculations at this $Q^2$, both of which converge to the GDH Sum value at $Q^2$=0. To see convergence to the GDH value in the data, it is clear that additional measurements at even lower $Q^2$ are needed.

The spin structure functions and their moments provide a unique mechanism for testing effective theories of QCD and better understanding nucleon structure. Jefferson Lab E08-027 collected transversely and longitudinally polarized proton data at very low $Q^2$ and with excellent statistical precision. The longitudinal results from this experiment, namely the spin structure function $g_1$ and related moments ($\Gamma_1$, $I_A$, and $\gamma_0$), are presented in this paper for the first time. These results reveal a small but interesting difference from other proton $g_1$ measurements in this kinematic region, and illustrate the importance of high-precision data near the pion production threshold. Our results provide a cross-check with the Hall B data on the magnitude of $\Gamma_1$ and $I_A$ at low $Q^2$ and consequently on the validity of the GDH slope and sum rule. 
These results provide another important piece of the nucleon structure puzzle which, along with other previously published results for the spin structure functions, helps to better illuminate the structure of our world's most prevalent particle.

\section*{Acknowledgments}
We thank the polarized target group, and the accelerator and Hall A staff for their work on E08-027. We would also like to thank Sebastian Kuhn, Alexandre Deur, Marco Ripani, Darren Upton, and the rest of the EG4 collaboration for the inclusion of their data on the plots in this publication for comparison, and for very useful discussions and suggestions. This work was supported by the Department of Energy (DOE) under grants DE-FG02-88ER40410 (funding the University of New Hampshire Nuclear Physics group, received by K.S.), DE-FG02-96ER40950 (funding the University of Virginia Polarized Target Group, received by D.D.) and DE-AC02-06CH11357 (funding the Argonne National Lab group, received by P.R.). The Southeastern Universities Research Association operates the Thomas Jefferson National Accelerator Facility for the DOE under contract DE-AC05-06OR23177.

\section*{Author Contributions}
D.R., K.S., and J.P.C. contributed equally to writing this paper. The E08-027 spokespeople are K.S., J.P.C., A.C., and D.C., and contributed to preparing the experiment to run. All authors contributed to the research as part of the E08-027 collaboration by analyzing the data, taking shifts running the experiment, or operating and maintaining the accelerator and experimental equipment. 

\bibliography{apssamp}

@article{KUHN20091,
title = {Spin structure of the nucleon—status and recent results},
journal = {Progress in Particle and Nuclear Physics},
volume = {63},
number = {1},
pages = {1-50},
year = {2009},
issn = {0146-6410},
doi = {https://doi.org/10.1016/j.ppnp.2009.02.001},
url = {https://www.sciencedirect.com/science/article/pii/S0146641009000040},
author = {S.E. Kuhn and J.-P. Chen and E. Leader}
}

@Article{Simula:2001iy,
     author    = "Simula, S. and Osipenko, M. and Ricco, G. and Taiuti, M.",
     title     = "{Leading and higher twists in the proton polarized
                  structure function  g1(p) at large Bjorken x}",
     journal   = "Phys. Rev.",
     volume    = "D65",
     year      = "2002",
     pages     = "034017",
     SLIFERCOMMENT ={eprint    = "hep-ph/0107036"},
     archivePrefix = "arXiv",
     doi       = "10.1103/PhysRevD.65.034017",
     SLACcitation  = "%%CITATION = HEP-PH/0107036;%%"
}

@phdthesis{druth_thesis,
      author         = "Ruth, D.",
      title          = "{A Strong-QCD Regime Measurement of the Proton’s Spin Structure}",
      school         = "University of New Hampshire",
      url            = "{https://www.osti.gov/biblio/1900611}",
      year           = "2022",
}

@article{Bernard,
      author         = "Bernard, V. and Kaiser, N. and Meissner, U.--G.",
      title          = "{Chiral Dynamics in Nucleons and Nuclei}",
      journal        = "Int.J.Mod.Phys.",
      volume         = "E4",
      pages          = "193-346",
      doi            = "10.1142/S0218301395000092",
      year           = "1995",
      eprint         = "hep-ph/9501384",
      archivePrefix  = "arXiv",
      primaryClass   = "hep-ph",
      reportNumber   = "CRN-95-3, TK-95-1",
      SLACcitation   = "%%CITATION = HEP-PH/9501384;%%",
}

@article{GDH,
  title = "{Exact Sum Rule for Nucleon Magnetic Moments}",
  author = {Drell, S. D. and Hearn, A. C.},
  journal = {Phys. Rev. Lett.},
  volume = {16},
  issue = {20},
  pages = {908--911},
  numpages = {0},
  year = {1966},
  publisher = {American Physical Society},
  doi = {10.1103/PhysRevLett.16.908},
  url = {https://link.aps.org/doi/10.1103/PhysRevLett.16.908}
}

@article{pasechnik,
  title = {Nucleon spin structure at low momentum transfers},
  author = {Pasechnik, Roman S. and Soffer, Jacques and Teryaev, Oleg V.},
  journal = {Phys. Rev. D},
  volume = {82},
  issue = {7},
  pages = {076007},
  numpages = {7},
  year = {2010},
  month = {Oct},
  publisher = {American Physical Society},
  doi = {10.1103/PhysRevD.82.076007},
  url = {https://link.aps.org/doi/10.1103/PhysRevD.82.076007}
}

@article{burkert-ioffe,
title = {On the Q2 variation of spin-dependent deep-inelastic electron-proton scattering},
journal = {Physics Letters B},
volume = {296},
number = {1},
pages = {223-226},
year = {1992},
issn = {0370-2693},
doi = {https://doi.org/10.1016/0370-2693(92)90831-N},
url = {https://www.sciencedirect.com/science/article/pii/037026939290831N},
author = {V.D. Burkert and B.L. Ioffe}
}

@Article{MAID2007,
author="Drechsel, D.
and Kamalov, S. S.
and Tiator, L.",
title="{Unitary Isobar Model -- MAID2007}",
journal="The European Physical Journal A",
year="2007",
volume="34",
number="1",
pages="69",
issn="1434-601X",
doi="10.1140/epja/i2007-10490-6",
url="http://dx.doi.org/10.1140/epja/i2007-10490-6"
}

@article{g2p_nature,
      author         = "Ruth, D. and Zielinski, R. and Gu, C. and others",
      title          = "{Proton spin structure and generalized polarizabilities in the strong quantum chromodynamics regime}",
      journal        = "Nature Physics",
      year           = "2022",
      doi            = "10.1038/s41567-022-01781-y"
}

@article{Bosted3,
  title = {Empirical Fit to Precision Inclusive Electron--Proton Cross Sections in the Resonance region},
  author = {Christy, M. E. and Bosted, P. E.},
  journal = {Phys. Rev. C},
  volume = {81},
  issue = {5},
  pages = {055213},
  numpages = {8},
  year = {2010},
  publisher = {American Physical Society},
  doi = {10.1103/PhysRevC.81.055213},
  url = {http://link.aps.org/doi/10.1103/PhysRevC.81.055213}
}

@article{EG4_longpaper,
  title = {Measurement of the nucleon spin structure functions for $0.01 < {Q}^{2} < 1\phantom{\rule{0.16em}{0ex}}{\mathrm{GeV}}^{2}$ using CLAS},
  author = {Deur, A. and Kuhn, S. E. and others},
  collaboration = {The CLAS Collaboration},
  journal = {Phys. Rev. C},
  volume = {111},
  issue = {3},
  pages = {035202},
  numpages = {34},
  year = {2025},
  month = {Mar},
  publisher = {American Physical Society},
  doi = {10.1103/PhysRevC.111.035202},
  url = {https://link.aps.org/doi/10.1103/PhysRevC.111.035202}
}

@article{MAMI_gamma0,
author = {Gurevich, G.M. and Lisin, Valery},
year = {2017},
month = {01},
pages = {111-116},
title = {Measurement of the proton spin polarizabilities at MAMI},
volume = {48},
journal = {Physics of Particles and Nuclei},
doi = {10.1134/S1063779617010105}
}

@article{EG1b2,
      author         = "Fersch, Robert and others",
      title          = "{Determination of the Proton Spin Structure Functions for
                        $0.05 < Q^2 < 5$ GeV$^2$ using CLAS}",
      collaboration = {CLAS Collaboration},
      journal = {Phys. Rev. C},
      volume = {96},
      issue = {6},
      pages = {065208},
      numpages = {34},
      year = {2017},
      month = {Dec},
      publisher = {American Physical Society},
      doi = {10.1103/PhysRevC.96.065208},
      url = {https://link.aps.org/doi/10.1103/PhysRevC.96.065208}
}

@article{Gold,
  title = {Forward doubly-virtual Compton scattering off the nucleon in chiral perturbation theory. II. Spin polarizabilities and moments of polarized structure functions},
  author = {Alarc\'on, Jose Manuel and Hagelstein, Franziska and Lensky, Vadim and Pascalutsa, Vladimir},
  journal = {Phys. Rev. D},
  volume = {102},
  issue = {11},
  pages = {114026},
  numpages = {26},
  year = {2020},
  month = {Dec},
  publisher = {American Physical Society},
  doi = {10.1103/PhysRevD.102.114026},
  url = {https://link.aps.org/doi/10.1103/PhysRevD.102.114026}
}

@article{EG1b,
title = "{Moments of the spin structure functions and for $g_1^p$ and $g_1^d$ for 0.05 $<$ $Q^2$ $<$ 3.0 GeV$^2$}",
journal = "Physics Letters B ",
volume = "672",
number = "1",
pages = "12 - 16",
year = "2009",
note = "",
issn = "0370-2693",
doi = "http://dx.doi.org/10.1016/j.physletb.2008.12.063",
url = "http://www.sciencedirect.com/science/article/pii/S0370269309000033",
author = "Y. Prok and P. Bosted and V.D. Burkert and A. Deur and K.V. Dharmawardane and G.E. Dodge and K.A. Griffioen and S.E. Kuhn and R. Minehart and others"
}

@article{Krebs,
  title = "{New Insights into the Spin Structure of the Nucleon}",
  author = {Bernard, V. and Epelbaum, E. and Krebs, H. and Meissner, U.--G.},
  journal = {Phys. Rev. D},
  volume = {87},
  issue = {5},
  pages = {054032},
  numpages = {9},
  year = {2013},
  publisher = {American Physical Society},
  doi = {10.1103/PhysRevD.87.054032},
  url = {http://link.aps.org/doi/10.1103/PhysRevD.87.054032}
}

@article{CHEN_2010,
   title={MOMENTS OF SPIN STRUCTURE FUNCTIONS: SUM RULES AND POLARIZABILITIES},
   volume={19},
   ISSN={1793-6608},
   url={http://dx.doi.org/10.1142/S0218301310016405},
   DOI={10.1142/s0218301310016405},
   number={10},
   journal={International Journal of Modern Physics E},
   publisher={World Scientific Pub Co Pte Lt},
   author={Chen, J.-P.},
   year={2010},
   month=oct, pages={1893–1921} }

@article{EG4_final,
      author         = "Zheng, X. and Deur, A. and Kang, H. and others",
      title          = "{Measurement of the proton spin structure at long distances}",
      journal        = "Nature Physics",
      volume         = "17",
      year           = "2021",
      pages          = "736-741",
      doi            = "10.1038/s41567-021-01198-z"
}

\end{document}